\documentclass[preprint, amsmath,amssymb, superscriptaddress]{revtex4-1}

\usepackage{graphicx}
\usepackage{xcolor,color}

\begin{document}

\title{Kink in cuprates: the role of the low-energy density of states}
\author{E.\,Razzoli}
\thanks{Equally contributed author}
\email[]{elia.razzoli@psi.ch}
\affiliation{SwissFEL, Paul Scherrer Institut, 5232 Villigen PSI, Switzerland}
\affiliation{Quantum Matter Institute, University of British Columbia, Vancouver, BC V6T 1Z4, Canada}
\author{F.\,Boschini}
\thanks{Equally contributed author}
\email[]{fabio.boschini@inrs.ca}
\affiliation{Quantum Matter Institute, University of British Columbia, Vancouver, BC V6T 1Z4, Canada}
\affiliation{Centre \'{E}nergie Mat\'{e}riaux T\'{e}l\'{e}communications, Institut National de la Recherche Scientifique, Varennes, Qu\'{e}bec J3X 1S2, Canada}
\author{M.\,Zonno}
\affiliation{Canadian Light Source, Inc. 44 Innovation Boulevard, Saskatoon, SK S7N 2V3, Canada}
\affiliation{Quantum Matter Institute, University of British Columbia, Vancouver, BC V6T 1Z4, Canada}
\author{M.\,X.\,Na}
\affiliation{Quantum Matter Institute, University of British Columbia, Vancouver, BC V6T 1Z4, Canada}
\affiliation{Department of Physics $\&$ Astronomy, University of British Columbia, Vancouver, BC V6T 1Z1, Canada}
\author{M.\,Michiardi}
\affiliation{Quantum Matter Institute, University of British Columbia, Vancouver, BC V6T 1Z4, Canada}
\affiliation{Department of Physics $\&$ Astronomy, University of British Columbia, Vancouver, BC V6T 1Z1, Canada}
\affiliation{Max Planck Institute for Chemical Physics of Solids, Nöthnitzer Straße 40, Dresden 01187, Germany}
\author{M.\,Schneider}
\affiliation{Quantum Matter Institute, University of British Columbia, Vancouver, BC V6T 1Z4, Canada}
\affiliation{Department of Physics $\&$ Astronomy, University of British Columbia, Vancouver, BC V6T 1Z1, Canada}
\author{E.\,H.\,da Silva Neto}
\affiliation{Quantum Matter Institute, University of British Columbia, Vancouver, BC V6T 1Z4, Canada}
\affiliation{Department of Physics, Yale University, New Haven, Connecticut 06511, USA}
\affiliation{Energy Sciences Institute, Yale University, West Haven, Connecticut 06516, USA}
\affiliation{Department of Applied Physics, Yale University, New Haven, Connecticut 06516, USA}
\author{S.\,Gorovikov}
\affiliation{Canadian Light Source, Inc. 44 Innovation Boulevard, Saskatoon, SK S7N 2V3, Canada}
\author{R.\,D.\,Zhong}
\affiliation{Condensed Matter Physics and Materials Science, Brookhaven National Laboratory, Upton, NY 11973, USA}
\author{J.\,Schneeloch}
\affiliation{Condensed Matter Physics and Materials Science, Brookhaven National Laboratory, Upton, NY 11973, USA}
\affiliation{Department of Physics $\&$ Astronomy, Stony Brook University, Stony Brook, NY 11795-3800, USA}
\author{G.\,D.\,Gu}
\affiliation{Condensed Matter Physics and Materials Science, Brookhaven National Laboratory, Upton, NY 11973, USA}
\author{S.\,Zhdanovich}
\author{A.\,K.\,Mills}
\author{G.\,Levy}
\author{D.\,J.\,Jones}
\affiliation{Quantum Matter Institute, University of British Columbia, Vancouver, BC V6T 1Z4, Canada}
\affiliation{Department of Physics $\&$ Astronomy, University of British Columbia, Vancouver, BC V6T 1Z1, Canada}
\author{C.\,Giannetti}
\affiliation{Department of Mathematics and Physics, Universit\`{a} Cattolica del Sacro Cuore, Brescia, BS I-25121, Italy}
\affiliation{Interdisciplinary Laboratories for Advanced Materials Physics (ILAMP),Universit\`{a} Cattolica del Sacro Cuore, Brescia I-25121, Italy}
\author{A.\,Damascelli}
\email[]{damascelli@physics.ubc.ca}
\affiliation{Quantum Matter Institute, University of British Columbia, Vancouver, BC V6T 1Z4, Canada}
\affiliation{Department of Physics $\&$ Astronomy, University of British Columbia, Vancouver, BC V6T 1Z1, Canada}

\date{\today}

\maketitle
\textbf{The 40-70\,meV band-structure renormalization (so-called kink) in high-temperature cuprate superconductors -- which has been mainly interpreted in terms of electron-boson coupling -- is observed to be strongly suppressed both above the superconducting transition temperature and under optical excitation. We employ equilibrium and time-resolved angle-resolved photoemission spectroscopy, in combination with Migdal-Eliashberg simulations, to investigate the suppression of the near-nodal kink in Bi$_2$Sr$_2$CaCu$_2$O$_{8+\delta}$. We show that the $\sim$30$\%$ decrease of the kink strength across the superconducting-to-normal-state phase transition can be entirely accounted for by the filling of the superconducting gap, without additional consideration of temperature- and time-dependent electron-boson coupling. Our findings demonstrate that consideration of changes in the density of states is essential to quantitatively account for the band structure renormalization effects in cuprates.}

\section*{Introduction}
Cuprate high-temperature superconductors are an invaluable platform for the study of emergent phenomena driven by strong electron interactions. However, despite extensive experimental and theoretical efforts over the last three decades, it remains still unclear whether and how different scattering channels (\textit{e.g.}, electron-spin fluctuations, electron-phonon, electron-electron) bear a direct relation to the elusive high-temperature superconducting glue. 
To this day, a complete understanding of how the electronic band structure of cuprates is renormalized by electron-boson interactions across the superconducting-to-normal-state phase transition is still lacking. 

The precise estimate of the strength and evolution of the electron-boson coupling $\lambda$, along with its impact on the electronic band structure, is a formidable challenge. Although new methods for extracting $\lambda$ have been recently proposed \cite{na2019direct,5iwick2019graphite,Ghiringhelli2019electronphonon}, to date angle-resolved photoemission spectroscopy (ARPES) at equilibrium is still one of the most consistent techniques used to evaluate the coupling of specific electronic states to bosonic modes. By accessing directly the one-electron removal spectral function $A(\textbf{k},\omega)$ and the related complex many-body self-energy $\Sigma$=Re$\Sigma$+$i$Im$\Sigma$, a multitude of ARPES studies have investigated the renormalization of the band dispersion in the 40-70\,meV binding energy range (so-called \emph{kink}), which is ubiquitous in cuprates \cite{kinkLanzaraNature}. Electron-spin fluctuations, electron-phonon, and electron-electron interactions have all been proposed as possible origins for the kink in cuprates  \cite{kinkLanzaraNature,BorisenkoKinkYBCO,VishikHgCuprate,Iwasawa2008KinkIsotope,Devereaux2004anisotropicElePhon,Lee2007,RuizBadia2009TheoryKink,Mou2017resonantModeKink,Kordyuk2004KinkResMode,Chubukov2004KinkOrigin,Terashima2007ImpurityEffectsKink,KinkelectrOriginNatPhys2007,Li2021, Huang2021}. 
Regardless of which boson mediates the band renormalization, an enhancement of the kink strength upon cooling across $T_c$ has been reported in cuprates and has been interpreted as a signature of an enhanced $\lambda$ in the superconducting state \cite{Terashima2007ImpurityEffectsKink,Kordyuk2006NodalKink,Lee2008InterplayKink_SC,Mou2017resonantModeKink}, despite the fact that in BCS-like analyses of electron-boson coupling the Eliashberg function $\alpha^2 F(\nu)$ is usually considered a temperature independent quantity \cite{NatPhysDalConte, Dahm2009}.

Time-resolved ARPES (TR-ARPES) provides direct access with momentum resolution to light-induced electron dynamics and transient changes of many-body interactions in cuprates \cite{zonno2021time} and, generally, quantum materials \cite{boschini2023time}. In particular, a transient weakening of the kink strength in cuprates (with no variation in its energy position) has been reported via TR-ARPES upon a near-IR optical excitation \cite{NatCommLanzara}. This weakening, together with the concomitant quench of the superconducting gap, has been attributed to changes in the electron-boson coupling strength $\lambda$, and a breaking of the Migdal-Eliashberg electron-boson description has been suggested \cite{ScienceLanzara,NatCommLanzara}.
Recently, both equilibrium and time-resolved ARPES measurements have demonstrated the filling of the superconducting gap (instead of its closure) as a direct consequence of an increased pair-breaking scattering rate  \cite{TDOSNatPhys,NatCommShin2015,DessauPRX,PerfettiFillingGap, Boschini2018,zonno2021time}. In light of these new findings, it is critical to revisit the role of the superconducting gap -- as well as of the effect of the near-IR photoexcitation -- on the kink in superconducting cuprates.

In this article, we provide a comprehensive description of the temperature dependence of the near-nodal kink, whose analysis is not complicated by pseudogap contributions. By combining spectral-function simulations within the Migdal-Eliashberg theory framework with equilibrium and TR-ARPES measurements on Bi$_2$Sr$_2$CaCu$_2$O$_{8+\delta}$ (Bi2212; underdoped T$_c$=82\,K, UD82; and optimally-doped T$_c$=91\,K, OP91), we show that the pump- and temperature-induced suppression of the kink strength can be explained purely in terms of a modification of the density of states following the filling of the superconducting gap, while consideration of a change in electron-boson coupling is not needed. Our findings imply that bosonic scattering vectors connect the node to states along the whole gapped Fermi surface contour, up to the antinodal region. 

\begin{figure*}[t]
\centering
\includegraphics[width=\textwidth]{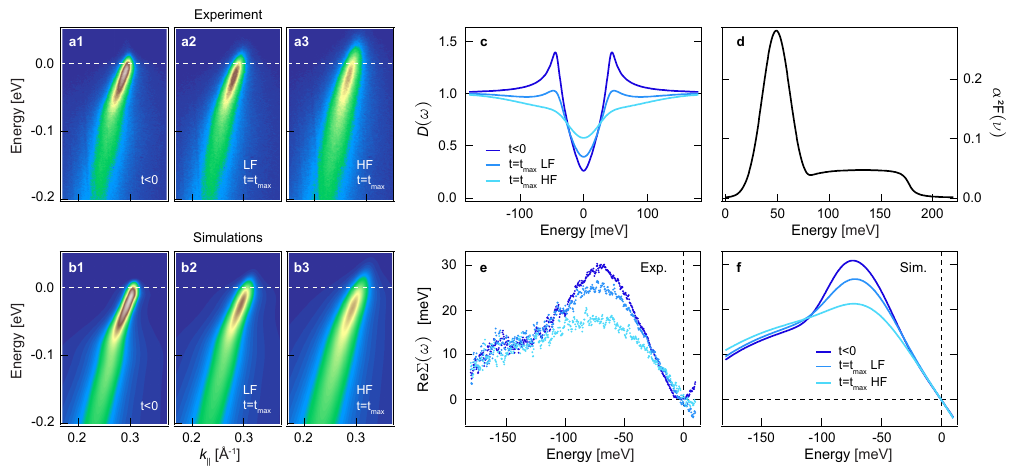}
\caption{ARPES data and simulations of the nodal spectral function and self-energy in Bi2212-UD82. \textbf{a1}-\textbf{a3} ARPES intensity maps along the nodal direction for low fluence (LF) and high fluence (HF) pump excitation; $t=t_{\text{max}}\approx$0.6\,ps is the pump-probe delay at which variation in the kink renormalization is maximized. \textbf{b1}-\textbf{b3} Simulated ARPES maps using Eqs.\,\ref{EQ1},\ref{EQ2}; convolutions to account for the energy and momentum experimental resolutions are applied. \textbf{c} Spectral weight density $D(\omega)$ at the Fermi momentum ($k_F$) integrated over the full Fermi surface; a $d$-wave superconducting gap with an antinodal value of 43\,meV was used. \textbf{d} $\alpha^2 F(\nu)$ used for the simulation; the functional form is adjusted in order to reproduce the experimental Re$\Sigma$ for $t<0$;  $\alpha^2 F(\nu)$ is considered both temperature- and time-independent, in agreement with previous works \cite{Cilento2013, vanHeumen2009}. \textbf{e} Real part of the self-energy extracted via momentum distribution curves (MDCs) fitting of the ARPES data shown in panels a1-a3; a linear bare dispersion has been assumed. \textbf{f} Real part of the nodal self-energy simulated using Eq.\,\ref{EQ1}, and $D(\omega)$ and $\alpha^2 F(\nu)$ shown in c and d. Electronic temperatures $T_e$ have been estimated via Fermi-Dirac fitting, and determined to be 20\,K and 70\,K for LF at $t<0$ and $t=t_{\text{max}}$, respectively, and 130\,K for HF at $t=t_{\text{max}}.$}
\label{Fig1}
\end{figure*}

\section*{Isotropic electron-boson model} 
Our simulations evaluate the nodal self-energy in the framework of Migdal-Eliashberg theory with isotropic electron-boson coupling \cite{Mahan2000,scalapino1966strong,pavarini2013emergent}:
\begin{align}
\Sigma_E (\omega) = \int_0^{\infty}  \int_{-\infty}^{\infty} d \nu d\omega' D(\omega')   \alpha^2 F(\nu)  K_T(\omega, \omega', \nu ),
\label{EQ1}
\end{align}
where $K_T$ is the $T$-dependent electron-boson kernel, $\alpha^2 F(\nu)$ is the frequency spectrum of the bosons coupling to the electrons, and $D(\omega) $ is the low-energy density of states. In our simplified model, $D(\omega)$ is calculated using the self-energy proposed by Norman et al., $\Sigma_N = -i \Gamma_s + \Delta^2 / (\omega + i \Gamma_p)$ \cite{NormanSelfEnergy} (and it is integrated over the full Fermi surface, see Supplemental Information for more details):
\begin{equation}
D^N(\omega) =   A_{\text{coh}} \pi \int d \theta'_{FS}    Re  [ \frac{\omega + i \gamma }{\sqrt{(\omega + i \gamma)^2-\Delta_{\theta'_{FS}}^2}} ],
\label{EQ2}
\end{equation}
where $\Gamma_s$ ($\Gamma_p$) is the single particle (pair-breaking) scattering rate, $\gamma=(\Gamma_s+\Gamma_p) / 2$, $\Delta_{\theta'_{FS}}$ is the $d$-wave superconducting gap (43\,meV at the antinode), and $A_{\text{coh}}$ is the coherent spectral weight amplitude at k$_F$. Note that models alternative to the one of Norman et al.\,\cite{NormanSelfEnergy}, such as Refs.\,\onlinecite{Herman2016, Herman2017}, can generally be used to evaluate $D(\omega)$.

\section*{Results} 
\subsection*{Light-induced quench of the nodal kink}
Figure\,\ref{Fig1}a1--a3 display the band mapping of Bi2212-UD82 along the nodal direction, for different excitation fluences. While panel (a1) shows the nodal ARPES spectrum before the pump arrival ($t<0$), panels (a2) and (a3) present spectra acquired at $t=t_{\text{max}}\approx$0.6\,ps for low-fluence (LF) and high-fluence (HF) regimes, respectively (defined in the Methods); $t_{\text{max}}$ identifies the time delay where the largest modification of the kink is observed, and corresponds to the maximal filling of the superconducting gap \cite{Boschini2018,ScienceLanzara,PerfettiFillingGap,zonno2021time}.
Figure\,\ref{Fig1}b1--b3 display simulated nodal ARPES spectra, which well reproduce their experimental counterparts in panels (a1)-(a3). 

Near-IR optical excitation prompts a transient increase and evolution of the electronic temperature $T_e(t)$, as well as of the single-particle and pair-breaking scattering terms, which reflect into the dynamical evolution of $\gamma (t)$, $\Delta (t)$, and $A_{\text{coh}} (t)$. Consequently, $K_{T_e}(\omega, \omega', \nu)$ and $D^N(\omega,t)$ acquire a time-dependence.
The time dependence of all terms contributing to $D^N(\omega,t)$, displayed in Fig.\,\ref{Fig1}c, has been extracted from nodal and near-nodal ($\Delta \approx$15\,meV) TR-ARPES data via a global fit of the experimental energy and momentum distribution curves (EDCs and MDCs)  \cite{Boschini2018, SM,zonno2021ubiquitous}. The loss of coherent spectral weight, which has been ubiquitously reported along the nodal direction and at the antinode as a function of the pump excitation as well as temperature \cite{fedorov1999temperature, feng2000signature, ding2001coherent,Graf2011,zonno2021ubiquitous}, is taken into account by reducing the coherent low-energy part of $D^N(\omega,t)$  by the corresponding factor $A_{\text{coh}}$. In contrast to previous works, we take $\alpha^2F$ to be both \emph{temperature-} and \emph{time-independent} (see Fig.\,\ref{Fig1}d); its functional form has been adjusted to reproduce the experimental Re$\Sigma$ for $t<0$. The comparison between calculated and experimental Re$\Sigma$ is in very close agreement, as shown in Fig.\,\ref{Fig1}e and f. 

By visual inspection of Fig.\,\ref{Fig1}e and f, the peak position of both the experimental and calculated Re$\Sigma$ is unchanged upon pump excitation, even at HF. 
Under the assumption of an \emph{isotropic electron-boson model} (\emph{i.e.}, gapless nodal states are connected to all other states in momentum space via coupling with a boson of energy  $\Omega_b$), the peak position of Re$\Sigma$ would fall approximately at  $\Omega_b + \langle \Delta \rangle_{\text{FS}} $, where $\langle \Delta \rangle_{\text{FS}}$ is the superconducting gap averaged over the entire Fermi surface. The gap amplitude at $t_{\text{max}}$  changes only of few percent even for the highest excitation fluence ($\Delta_{AN}= 43$, $41$, and $35$\,meV for $t<0$ and at $t_{\text{max}}$ for LF and HF, respectively, as extrapolated by off-nodal TR-ARPES data \cite{Boschini2018,zonno2021time}). Instead, the filling of $D^N(\omega)$ [see Fig.\,\ref{Fig1}c] is governed primarily by the pair-breaking scattering rate [$\Gamma_p^{\text{LF}} (t_{\text{max}})=8$\,meV, $ \Gamma^{\text{HF}}_p(t_{\text{max}})=25$\,meV]. We remark that our model also captures the kink evolution for off-nodal states (see Supplemental Information), further emphasizing that modification of the density of states is responsible for the observed reduction of the kink renormalization.

\begin{figure}[t]
\includegraphics[scale=1]{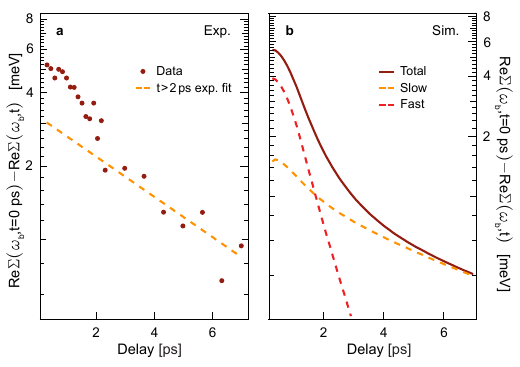}
\caption[Fig3]{Dynamics of the nodal kink from the ARPES data and corresponding simulations. \textbf{a} Re$\Sigma(t)$ at the kink energy ($\omega_\text{b}\sim$70\,meV) obtained from MDC fits of ARPES data. The line is an exponential fit of the data for $t>2$ ps (decay time $\tau = 4.9$ ps).  \textbf{b} Simulation of dynamical change of Re$\Sigma$ (burgundy line) using $\alpha^2F$ of Fig.\,\ref{Fig1}d, which is temperature and time-independent; orange and red lines are the contribution from slow ($\Delta$, $\Gamma_s$ and $T_e$ in $D^N$ and $K$) and fast ($\Gamma_p$ and $A_{coh}$ in $D^N$) decaying terms, respectively. }
\label{Fig3}
\end{figure}

\subsection*{Ultrafast evolution of the nodal kink}
Having established the mechanism behind the change in the kink at fixed time delays, we now discuss the dynamics of the kink in relation to pair-breaking scattering events. Figure\,\ref{Fig3}a shows the time evolution of Re$\Sigma$ at the kink energy for $t >0$, in logarithmic scale. 
We observe two distinct contributions, namely a fast ($\tau <2$\,ps) and a slow ($\tau \sim 4.9$\,ps) decay. The response at the fast timescale is mostly ascribed to the pump-induced enhancement of phase fluctuations encoded in $\Gamma_p$ and the suppression of the coherent spectral weight $A_{coh}$. Instead, the slow decay term has a thermal origin and follows the evolution of $\Delta$, $\Gamma_s$ and  $T_e$.
Figure\,\ref{Fig3}b displays the calculated time evolution of the kink according to the experimentally extracted dynamics of the parameters of the model of Eq.\,\ref{EQ1} (see Supplemental Information). Contributions to the total Re$\Sigma$ dynamics from slow ($\Delta$, $\Gamma_s$ and $T_e$) and fast ($\Gamma_p$ and $A_{coh}$) decaying terms are shown in Fig.\,\ref{Fig3}b as orange and red dashed lines, respectively. Together, they reproduce well the observed timescales and dynamics of Re$\Sigma$ extracted from the data (burgundy line), thereby confirming that the light-induced enhancement of pair-breaking scattering rate underlies the decrease of the kink strength within the first 2\,ps.

\begin{figure}[t]
\includegraphics[scale=1]{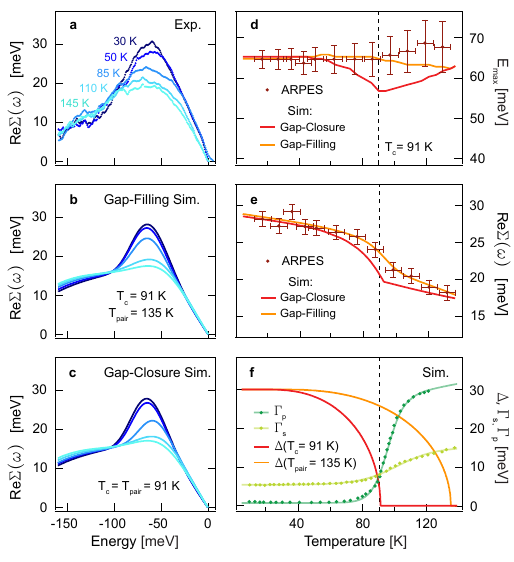}
\caption[Fig4]{ARPES data and simulations of T-dependence of Re$\Sigma$ for Bi2212-OP91. \textbf{a} Re$\Sigma$ extracted via MDC fitting of the ARPES data. \textbf{b} and \textbf{c} Simulated Re$\Sigma$ for the gap-filling and gap-closure scenarios, respectively. \textbf{d} Temperature dependence of kink energy from ARPES data and simulations.  \textbf{e} Temperature dependence of Re$\Sigma$ maximal amplitude from ARPES data and simulations. \textbf{f} T-dependence of the parameters in the model (adapted from Ref.\,\onlinecite{Kondo2015}); although the superconductive condensate vanish at $T_c$ for both gap-closure and gap-filling scenarios, the superconducting gap closes at $T_c$=91\,K and $T_{\text{pair}}$=135\,K, respectively.}
\label{Fig4}
\end{figure}

\subsection*{Temperature evolution of the nodal kink via equilibrium ARPES}
In pump-probe experiments, pair-breaking events arise from the interaction between the superconducting condensate and the pump-induced nonthermal population of bosons \cite{Boschini2018}. However, in the following, we demonstrate that our isotropic electron-boson model with a \emph{temperature-independent} $\alpha^2F$ also successfully reproduces the static temperature dependence of the kink across the superconducting-to-normal-state phase transition. 
Figure\,\ref{Fig4}a displays the temperature dependence of Re$\Sigma$  extracted from our equilibrium ARPES data on Bi2212-OP91. Isotropic electron-boson model calculations for gap-filling and gap-closure scenarios are shown in panels (b) and (c), respectively.
In contrast to the gap-closing scenario where the superconducting gap closes at $T_c$, in the gap-filling scenario the superconducting-to-normal state phase transition is driven by the loss of macroscopic phase coherence at $T_c$ while the superconducting gap closes at $T_{\text{pair}}>T_c$ \cite{Kondo2015,TDOSNatPhys}.
The $T$-dependence of scattering rates and superconducting gap used for the simulations of Fig.\,\ref{Fig4}b,c are taken from Ref.\,\onlinecite{Kondo2015} and displayed in Fig.\,\ref{Fig4}f. In the gap-filling scenario, $T_{\text{pair}} \approx$135\,K for Bi2212-OP91 and $A_{coh}$ is assumed to decrease with $T$, in agreement with our measurements and previous works \cite{Graf2011,zonno2021ubiquitous}.

The gap-filling scenario captures well the absence of the shift in the kink energy at $T_c$ [see Fig.\,\ref{Fig4}d, orange line] and reproduces the change in the peak intensity of Re$\Sigma$ for the whole temperature range, as shown in Fig.\,\ref{Fig4}e, orange line. Instead, in the gap-closure scenario, the peak position of Re$\Sigma$ clearly shifts as a function of the temperature [red line in Fig.\,\ref{Fig4}d], in disagreement with our and previous experiments \cite{kinkLanzaraNature,BorisenkoKinkYBCO,VishikHgCuprate}. Moreover, simulations with gap closure at $T_c$ overestimate the change of the Re$\Sigma$ at the kink energy across $T_c$ [Fig.\,\ref{Fig4}e, red line], further highlighting the shortcomings of this approach \cite{Sandvik2004}.

\section*{Discussion}
Previously proposed models interpreted the lack of energy shift in the energy position of the kink across $T_c$ as evidence for strong momentum dependence of electron-boson interaction \cite{Kulic2005}. In that small-$q$ forward scattering picture, electron scattering occurs only within a small volume in momentum space, and the kink energy is given by $\Omega_b+\Delta_{\theta'_{FS}}$. While this approach explains the lack of energy shift at the node, where $\Delta_{\theta'_{FS}} = 0$ above and below $T_c$, it also predicts an increase in the kink energy moving away from the node -- due to the increase in $\Delta_{\theta'_{FS}}$ -- which is not observed experimentally \cite{He2013LaserKink}. Our model, although using a rather simplistic isotropic electron-boson interaction with a momentum-, temperature-, and time-independent $\alpha^2F$, well captures both the temperature dependence and dynamical response of the kink self-energy in the near-nodal region upon the filling of the superconducting gap.
In addition, in contrast with the small-$q$ forward scattering picture, our model does not predict an increase in the kink energy moving away from the nodal direction since it is only dependent on the averaged gap $\langle \Delta \rangle_{\text{FS}}$ along the Fermi surface (see Supplemental Information for more details).

In summary, our work provides a comprehensive understanding of the band renormalization (kink) in Bi-based cuprates by combining spectral-function simulations within the Migdal-Eliashberg theory framework with equilibrium and out-of-equilibrium ARPES measurements. We demonstrated that an isotropic model for the electron-boson coupling thoroughly describes the phenomenology of the near-nodal kink in the cuprates when changes in the density of states are taken into account. 
Our model captures quantitatively both the photoinduced sub-ps dynamics as well as the equilibrium temperature dependence of the kink strength and energy position without the need to invoke an ad-hoc enhancement of coupling or the breakdown of the Migdal-Eliashberg theory.

\section*{Methods\label{Methods}}
\textbf{TR-ARPES and ARPES measurements}\\TR-ARPES data were acquired at the UBC-Moore Center for Ultrafast Quantum Matter using 1.55\,eV pulses for optical excitation (pump) and $\sigma$-polarized 6.2\,eV pulses for photoemission (probe). The overall energy, momentum, and temporal resolution of the system were 18\,meV, 0.0025\,$\text{\AA}^{-1}$ and 250\,fs, respectively \cite{Boschini2018}. Two different fluence regimes were explored for TR-ARPES experiments: low fluence $\sim$8\,$\mu$J/cm$^2$ (LF), and high fluence $\sim$30\,$\mu$J/cm$^2$ (HF). The HF regime fully quenches the macroscopic superconducting condensate \cite{ScienceLanzara,Boschini2018}.
Equilibrium ARPES measurements were performed at the Quantum Materials Spectroscopy Centre (QMSC) beamline at the Canadian Light Source, using 27\,eV $\sigma$-polarized light, with energy and momentum resolutions better than 4.5\,meV and 0.007\,$\text{\AA}^{-1}$, respectively.

\textbf{Samples}\\ Single-crystal Bi$_2$Sr$_2$CaCu$_2$O$_{8+\delta}$ (Bi2212; underdoped T$_c$=82\,K, UD82; and optimally-doped T$_c$=91\,K, OP91) samples have been grown using the floating-zone method and hole-doped by oxygen annealing.
Bi2212 samples were aligned via Laue diffraction along the nodal $\overline{\Gamma}$--$\overline{\text{Y}}$ direction in order to avoid replica bands \cite{king2011structural, saini1997topology, yang1998crystal}.
The samples were cleaved at $<$5$\cdot$10$^{-11}$ torr base pressure.

\textbf{Data Availability}\\
The data that support the plots within this paper and other findings of this study are available from the corresponding author upon reasonable request.

\section*{References\label{References}}
\bibliographystyle{apsrev4-1}
\providecommand{\noopsort}[1]{}\providecommand{\singleletter}[1]{#1}

\vspace{0.7 cm}

\begin{acknowledgments}
This project was undertaken thanks in part to funding from the Max Planck-UBC-UTokyo Center for Quantum Materials and the Canada First Research Excellence Fund, Quantum Materials and Future Technologies Program. This effort was also funded by the Gordon and Betty Moore Foundation’s EPiQS Initiative, Grant GMBF4779 to A.D.; the Killam, Alfred P. Sloan, and Natural Sciences and Engineering Research Council of Canada’s (NSERC's) Steacie Memorial Fellowships (A.D.); the Alexander von Humboldt Foundation (A.D.); the Canada Research Chairs Program (A.D.); NSERC; the Department of National Defence (DND); Canada Foundation for Innovation (CFI); the British Columbia Knowledge Development Fund (BCKDF); and the CIFAR Quantum Materials Program (A.D.). F.B. acknowledges support from NSERC, the Fonds de recherche du Qu\'{e}bec – Nature et Technologies (FRQNT), and the Minist\`{e}re de l'\'{E}conomie, de l'Innovation et de l'\'{E}nergie - Qu\'{e}bec.
Part of the research described in this work was performed at the Canadian Light Source, a national research facility of the University of Saskatchewan, which is supported by CFI, NSERC, the National Research Council (NRC), the Canadian Institutes of Health Research (CIHR), the Government of Saskatchewan, and the University of Saskatchewan. 
E.R. acknowledges support from the Swiss National Science Foundation (SNSF) grant no. P300P2-164649.
E.H.d.S.N. acknowledges prior support from the Max Planck-UBC postdoctoral fellowship and CIFAR Global Scholar program, and current support from the Alfred P. Sloan Fellowship in Physics and the National Science Foundation under grant no. DMR 2034345.
The work at BNL was supported by the US Department of Energy, oﬃce of Basic Energy Sciences, contract no. DOE-sc0012704.
\end{acknowledgments}

\clearpage
\newpage
\onecolumngrid
\renewcommand{\thefigure}{S\arabic{figure}}
\renewcommand{\theequation}{S\arabic{equation}}
\setcounter{figure}{0}
\setcounter{equation}{0} 
\section*{Supplementary Information}

\subsection{Migdal-Eliashberg theory}
\begin{figure*}[b]
\centering
\includegraphics[scale=1.25]{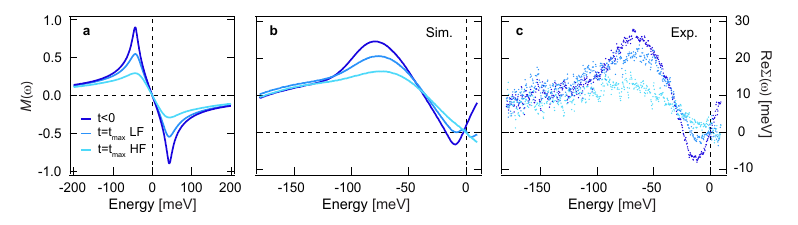}
\caption[FigS1]{ARPES data and simulations of the off-nodal dispersion in Bi2212-UD82. \textbf{a} Off-diagonal tomographic density of states integrated over the entire Fermi surface.
\textbf{b} Simulated Re$\Sigma$ extracted via MDC fitting of the intensity maps computed including both diagonal and off-diagonal spectral weight densities.  
\textbf{c} Re$\Sigma$ extracted via MDC fitting of the ARPES data. A linear bare dispersion has been assumed for both panels (b) and (c). Curves for $t<0$ and $t=t_{max}$ LF and HF are drawn in cyan, black and red, respectively.}
\label{S1}
\end{figure*}
The Green's functions in the Gorkov-Nambu formalism is
\begin{align}
\overline{\mathcal{G}} =
 \left(
\begin{array}{cc}
\mathcal{G}(k, \omega) & \mathcal{F}(k,\omega)  \\
\mathcal{F}(k, \omega) & -\mathcal{G}(k,-\omega)  \\
\end{array} \right)
= \frac{\omega Z(k,\omega)\tau_0 + [\epsilon_k+\chi(k,\omega)]\tau_3-\Phi(k,\omega)\tau_1}{[\omega Z(k,\omega)]^2-[\epsilon_k+\chi(k,\omega)]^2-\Phi(k,\omega)^2},
\label{EQ3}
\end{align}
and the two self-energies, within the Migdal-Eliashberg theory, are
\begin{align}
\Sigma_E(k, \omega) = \omega [1 - Z(k, \omega)] &=  \int_0^{\infty}  \int_{-\infty}^{\infty} d \nu d\omega' D(\omega')   \alpha^2 F(\nu)  K(\omega, \omega', \nu ),
\\
\Phi_E(k, \omega) &=   \int_0^{\infty}  \int_{-\infty}^{\infty} d \nu d\omega' M(\omega')   \alpha^2 F(\nu)  K(\omega, \omega', \nu ),
\end{align}
with 
\begin{align}
D(\omega) &=   N(0) \pi \int d \theta'_{FS}    Re  [ \frac{\omega Z(k,\omega) }{\sqrt{[\omega Z(k,\omega)]^2-\Phi(k,\omega)^2}} ],\\
M(\omega) &=   N(0) \pi \int d \theta'_{FS}    Re  [ \frac{\Phi(k,\omega) }{\sqrt{[\omega Z(k,\omega)]^2-\Phi(k,\omega)^2}} ],\\
K(\omega, \omega', \nu ) &= \frac{n(\nu) + 1 - f(\omega')}{\omega -\nu -\omega' + i \delta} + \frac{n(\nu) + f(\omega')}{\omega +\nu -\omega' + i \delta}. 
\end{align}
In our approximation, we substitute the values of $Z$ and $\Phi$ obtained from the phenomenological model of Norman et al. \cite{NormanSelfEnergy} in the expression for the diagonal ($D$) and off-diagonal ($M$) tomographic density of states integrated over the entire Fermi surface. In particular, we define $Z= 1 + i \gamma/\omega$, $\chi=i(\Gamma_P-\Gamma_s)/2=i \gamma'$, $\phi=\Delta$.
The model of Norman and co-authors is already a good guess of the self-consistent solution of the Migdal Eliashberg model, and it captures the filling of the superconducting gap following the increase of the pair-breaking scattering rate $\Gamma_P$.
Our model estimates variations of the electron dispersion induced by a change in the diagonal and off-diagonal density of states ($D$ and $M$) upon enhancement of $\Gamma_P$. 
Simulated intensity maps are thus obtained by plotting Im$\mathcal{G}(k, \omega)$ from Eq.\,(\ref{EQ3}), and the self energies are given by
\begin{align}
\overline{Z}(k, \omega, t)&=
1 +i \frac{\gamma}{\omega}- \frac{N(0) \pi}{\omega} \int_0^{\infty}  \int_{-\infty}^{\infty} d \nu d\omega'  \int d \theta'_{FS}    Re  [ \frac{\omega' + i \gamma(t) }{\sqrt{(\omega' + i \gamma)^2-\Delta_{\theta'_{FS}}^2}(t)}   \alpha^2 F(\nu)  K_t(\omega, \omega', \nu ),
\\
\overline{\Phi}(k, \omega, t) &= 
\qquad \qquad   N(0) \pi \int_0^{\infty}  \int_{-\infty}^{\infty} d \nu d\omega' \int d \theta'_{FS}    Re  [ \frac{\Delta_{\theta'_{FS}}(t) }{\sqrt{(\omega' + i \gamma)^2-\Delta_{\theta'_{FS}}^2(t)}}   \alpha^2 F(\nu)  K_t(\omega, \omega', \nu ),   
\end{align}
where also $K_t$ depends on the delay $t$ through the electronic and bosonic temperature.

In an effort to discuss the kink behavior in the near-nodal region  (\emph{i.e.}, when $\Delta\neq0$), Eqs. (1)-(2) need to be complemented by the off-diagonal counterpart. 
Figure\,\ref{S1}(a) displays $M(\omega)$ for $t<0$ (cyan) and $t=t_{max}$ LF (black) and HF (red).
The corresponding off-diagonal self-energy is calculated as a double integral of $M$, $\alpha^2 F$ and the kernel $K$. Similar to what is observed at the node, the pump-induced kink suppression is reproduced by our simple model [see Fig.\,\ref{S1}(b,c)], further emphasizing that modification of low energy electron density of states prompts the change in the kink dispersion.  

\begin{figure*}[t]
\includegraphics[scale=0.93]{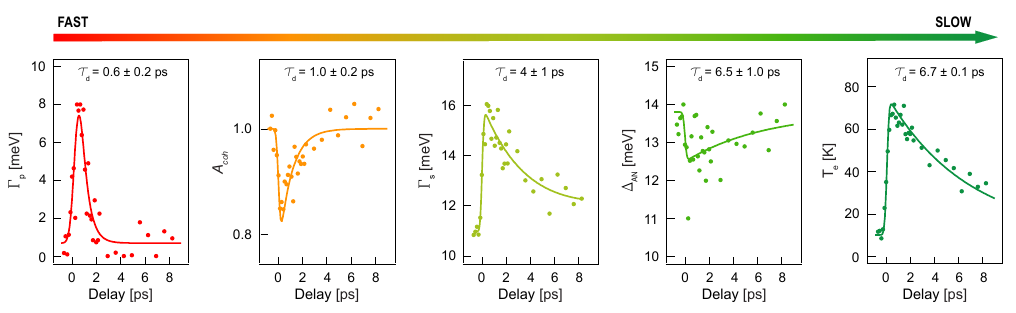}
\caption[S1]{Temporal evolution and single-exponential fit of the parameters used in our model.}
\label{S2}
\end{figure*}

\subsection{Time and temperature dependence of the parameters in the model}
The temporal dependence of the parameters used to produce the results of Fig.\,\ref{Fig3} is extracted from a single-exponential fit of the data shown in Fig.\,\ref{S2} [experimental data (symbols) are obtained according to the procedure introduced in Ref,\,\cite{Boschini2018}]. 

\begin{figure}[ht]
\includegraphics[scale=1.45]{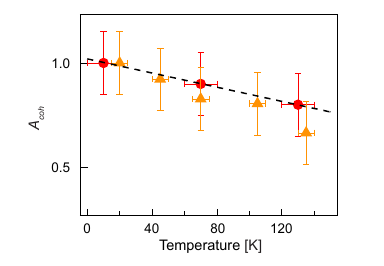}
\caption[S2]{Temperature dependence of $A_{coh}$. Red circles are the values from pump-probe data at the corresponding $T_e$, as described in Fig.\,\ref{Fig1}. Orange triangles are from our static ARPES data \cite{zonno2021ubiquitous}.   
The black dashed line is the linear dependence used for the calculations of Fig.\,\ref{Fig4}. }
\label{S3}
\end{figure}

The temperature dependence of the parameters used to produce the results of Fig.\,\ref{Fig4} are shown in Fig.\,\ref{Fig4}f and discussed in the main text except for $A_{coh}$, which is shown in Fig.\,\ref{S3}. Experimental data of $A_{coh}$ (symbols) are obtained by fitting the symmetrized EDC by a Voigt profile and calculating the subtended area.

\newpage

\end{document}